# Parameter Extraction and Support-Loss in MEMS Resonators

P.G. Steeneken[*,1], J.J.M. Ruigrok[1], S. Kang[1], J.T.M. van Beek[1], J. Bontemps[2] and J.J. Koning[2]
[1]NXP Semiconductors, Research, Eindhoven, The Netherlands
[2]NXP Semiconductors, Device Engineering & Characterization, Nijmegen, The Netherlands
*Corresponding author: HTC 4 (WAG02), 5656 AE Eindhoven, The Netherlands, peter.steeneken@nxp.com

**Abstract:** In this paper it is shown how the equivalent circuit parameters of a MEMS resonator can be simply obtained from an eigenfrequency simulation. Additionally, it is demonstrated that the *Q*-factor as a result of support losses in a MEMS resonator can be determined using a matched boundary layer. The method is applied to calculate the frequency dependent admittance of a diamond disk resonator. Results agree well with measurements and analytic results. Comparison to a frequency response analysis establishes the validity of the method and shows that it results in a large reduction of the simulation time.

## 1. Introduction

Because the *Q*-factors of mechanical resonators can be much higher than that of electrical resonators made from coils and capacitors, they are very interesting components for constructing oscillators and filters. However to use these mechanical resonator elements in an electrical circuit, an electromechanical transducer is needed to convert electrical energy to the mechanical domain.

Developments in microsystem technology have enabled the use of electrostatic actuation forces across sub-micron vacuum gaps to accomplish this energy conversion in micro-electromechanical systems (MEMS) resonators. Main advantages of these resonators are their integration in silicon, their small size and the possibility to control their resonance frequency by lithographic mask design.

Figure 1 shows a cross-section of a micromechanical disk resonator [1], which will be used as an exemplary MEMS resonator in this paper. The axially symmetric structure consists of a diamond disk with height *H* and outer radius *R*, which is fixed to a silicon substrate by a polysilicon stem with radius *a*. The acoustic mismatch between the diamond disk and the silicon stem increases the *Q*-factor of the resonator [1]. The disk is actuated by a radial electrostatic force which is generated by applying a voltage on the actuation electrode with area $A_{act}$, which is separated from the edge of the disk by a uniform gap *g*.

For making circuit designs using MEMS resonators, designers need an equivalent electrical circuit that describes their frequency dependent characteristics. An obvious way to determine the frequency dependent admittance $Y(\omega)$ of a MEMS resonator is using a frequency response analysis. In this type of analysis, an actuation force of a frequency $\omega$ is applied to the surface of the resonator and the amplitude response is simulated. Because the *Q*-factor of MEMS resonators is usually high, this method requires a high density of frequency points to resolve a single resonance of the resonator. Moreover, fitting of $Y(\omega)$ is required to extract the equivalent circuit parameters.

In this paper an alternative method is described to directly extract the circuit parameters from an eigenfrequency analysis. The method will be applied to the resonator in figure 1.

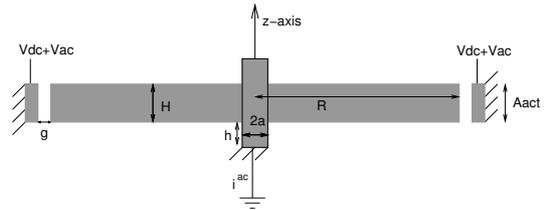

**Figure 1** Cross-section through the center of a diamond disk resonator, which is axially symmetric around the z-axis.

## 2. Single degree of freedom equations

Let us first consider a MEMS resonator system with only one mechanical degree of freedom *x*. Its equation of motion is identical to that of a drive harmonic oscillator:

$$F_{ac} e^{j\omega t} = m\frac{d^2 x(t)}{dt^2} + b\frac{dx(t)}{dt} + kx(t) \quad (1)$$

Assuming that the steady-state displacement can be written as:

$$x(t) = \operatorname{Re} x(\omega) e^{j\omega t} \quad (2)$$

Substitution in (1) gives:

$$x(\omega) = \frac{F_{ac}}{j\omega}\left[j\omega m + b + (k/j\omega)\right]^{-1} \quad (3)$$

The electrostatic force is given by the equation:

$$F_e = \frac{V^2}{2}\frac{dC(x)}{dx} \quad (4)$$

To be able to perform a linear analysis, it will be assumed that the maximum displacements are very small compared to the gap size *(x<<g)*, that the parallel plate approximation $C = A_{act}\varepsilon_0/(g-x)$ is valid *(g<<2πR and g<<H)* and that the dc bias voltage is much larger than the ac voltage $(V_{dc}>>V_{ac})$. The ac component of the electrostatic force as a result of a voltage $V(t)=V_{dc}+V_{ac}e^{j\omega t}$ is then given by:

$$F_{ac} = \frac{A_{act}\varepsilon_0 V_{dc}}{g^2}V_{ac} = \eta V_{ac} \quad (5)$$

From the equivalence between the mechanical power $P_{mech}=F_{ac}\times dx/dt$ and electrical power $P_e=V_{ac}\times i_{ac}$ the admittance $Y$ of the circuit can be calculated:

$$F_{ac}\frac{dx}{dt} = \eta V_{ac} j\omega x = V_{ac}i_{ac} = V_{ac}^2 Y \quad (6)$$

Substituting equations (3) and (5) in (6), we find that the admittance of the one-port small ac-signal MEMS resonator is given by:

$$Y(\omega) = \frac{\eta^2}{j\omega m + b + (k/j\omega)} \quad (7)$$

An equivalent electrical circuit with the same frequency dependent admittance as equation (7) is shown in figure 2 [2]. The ideal transformer with turns-ratio 1:$\eta$ represents the conversion from electrical to mechanical energy. The equivalent mass $m_i$, spring constant $k_i$ and damping $b_i$ can be represented by a coil, capacitor and resistor respectively. In the mechanical part of the circuit voltages are identical to forces and currents to speeds. The voltage at node $x$ is equivalent to the position $x$. Even if the resonator is not moving, a stray capacitance $C_w=A_{act}\varepsilon_0/g$ across the gap is present across the actuation gap.

Equation (7) was derived for a MEMS resonator with a single mechanical degree of freedom. A resonator with $N$ degrees of freedom will have $N$ eigenmodes. For each eigenmode $i$ of the resonator ($i=1,2,...,N$) a parameter set $m_i$, $k_i$ and $b_i$ exists. The complete equivalent circuit consists of these $N$ resonant circuits connected in parallel. Therefore the total frequency dependent admittance $Y(\omega)$ of a MEMS resonator with $N$ degrees of freedom can be written as:

$$Y(\omega) = \frac{i_{ac}}{V_{ac}} = j\omega C_w + \eta^2 \sum_{i=1}^{N}\left(j\omega m_i + b_i + \frac{k_i}{j\omega}\right)^{-1} \quad (8)$$

In the following sections it will be described how the parameters $m_i$, $k_i$ and $b_i$ can be simply obtained from the eigenvalue and the surface and volume integrals of the eigenmodes. This method is based on the normal-mode method of dynamic analysis [3].

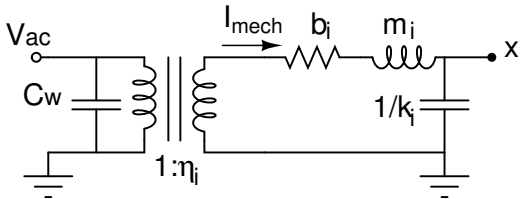

**Figure 2** MEMS resonator small ac signal equivalent electrical circuit.

## 3. Eigenfrequency analysis

We consider a resonator with many mechanical degrees of freedom. For modes with a high $Q$-factor, the mode-shapes do not depend much on the electrostatic force [2]. Therefore it is possible to extract $m_i$, $b_i$ and $k_i$ by solving the partial differential equation describing the elastic system for $F_{ac}=0$, which is called the translational equation of motion [4]:

$$\nabla\cdot(\mathbf{c}\nabla\mathbf{u}(\mathbf{r},t)) - \rho\frac{\partial^2\mathbf{u}(\mathbf{r},t)}{\partial^2 t} = 0 \quad (9)$$

Here **c** represents the stiffness tensor, $\rho$ is the mass density and $\mathbf{u}(\mathbf{r},t)$ is the particle displacement field. Solutions of equation (9) have a time dependence given by:

$$\mathbf{u}_i(\mathbf{r},t) = \mathrm{Re}(\mathbf{u}_i(\mathbf{r})e^{j\omega_i t}) \quad (10)$$

Substituting this in equation (9) we obtain the eigenvalue equation:

$$\nabla\cdot(\mathbf{c}\nabla\mathbf{u}_i(\mathbf{r})) + \rho\omega_i^2\mathbf{u}_i(\mathbf{r}) = 0 \quad (11)$$

The COMSOL finite element package is used to solve equation (11) in an eigenfrequency analysis, to obtain the eigenmodeshapes $\mathbf{u}_i(\mathbf{r})$ and the corresponding complex eigenfrequencies $\omega_i=2\pi f_i$.

## 4. Definition of the displacement $x_i$

To reduce this set of eigenmodes and eigenvalue solutions to a set of parameters $m_i$, $b_i$ and $k_i$, postprocessing of the solutions is required. First of all, a proper definition of the equivalent displacement parameter $x_i$ needs to be defined. To choose a proper definition, we note that the electrostatic work in the equivalent circuit should equal the work in the system with many degrees of freedom. Because work is given by the dot product between force and displacement vectors, the electrostatic work $W_e$ done by the electrostatic actuator equals the area integral over the dot product between electrostatic pressure and displacement $\mathbf{P}_e\cdot\mathbf{u}_i$ where $\mathbf{P}_e=\mathbf{F}_{ac}/A_{act}$.

$$W_e = F_{ac}x_i = \frac{1}{A_{act}}\int_{A_{act}}\mathbf{F}_{ac}\cdot\mathbf{u}_i\,dA \quad (12)$$

The area integral should be taken over the surface of the resonator that is facing an actuation electrode. In the parallel-plate approximation the electrostatic force is always perpendicular to the actuation surface. Therefore equation (12) is satisfied by defining the equivalent displacement $x_i$ as:

$$x_i \equiv \frac{1}{A_{act}}\int_{A_{act}}\mathbf{u}_i\cdot\mathbf{n}\,dA \quad (13)$$

Where **n** is the unit vector outward perpendicular to the surface.

The transduction factor $\eta$ is given by equation (5):

$$\eta = \frac{A_{act}\varepsilon_0 V_{dc}}{g^2} \qquad (14)$$

Note that this equation for $\eta$ does not depend on the eigenmode-shape. $\eta$ is therefore identical for all eigenmodes $i$.

## 5. Parameter extraction of $k_i$ and $m_i$

The stored energy $E_{stored}$ is a sum of the strain energy and the kinetic energy. The maximum kinetic and strain energy in the single degree of freedom system are given by $E_{kin,max}=½m_i|\omega_i x_i|^2$ and $E_{strain,max}=½k_i|x_i|^2$ and are both equal to $E_{stored}$. The solutions of $\mathbf{u}_i(\mathbf{r})$ of equation (11) obey:

$$\int_V W_s dV - \frac{1}{2}\int_V \rho\omega_i^2 \mathbf{u}_i \cdot \mathbf{u}_i dV = 0 \qquad (15)$$

The strain energy density is given by $W_s=½\varepsilon\cdot\sigma$, where $\varepsilon$ and $\sigma$ are the local strain and stress tensors. $V$ is the total simulated volume. The absolute values of the two energy integrals in equation (15) represent the stored energy $E_{stored}$. The equivalence between the stored energy in the single and many degree of freedom systems allows the extraction of the parameters $k_i$ and $m_i$ using the following equations:

$$\begin{aligned} E_{stored} &= \frac{1}{2}k_i|x_i|^2 = \left|\int_V W_s dV\right| \\ E_{stored} &= \frac{1}{2}m_i|\omega_i x_i|^2 = \frac{1}{2}\left|\int_V \rho\omega_i^2 \mathbf{u}_i \cdot \mathbf{u}_i dV\right| \\ k_i &= \frac{2}{|x_i|^2}\left|\int_V W_s dV\right| \\ m_i &= \frac{k_i}{|\omega_i|^2} = \frac{1}{|x_i|^2}\left|\int_V \rho\mathbf{u}_i \cdot \mathbf{u}_i dV\right| \end{aligned} \qquad (16)$$

Note that since $x_i$ should be calculated from (13), the values of $k_i$ and $m_i$ will depend on the location of the actuation electrodes.

## 6. The Q-factor in MEMS resonators

Calculation of the damping parameter $b_i$ is directly related to the calculation of the $Q$-factor by the equation:

$$b_i = \frac{\sqrt{k_i m_i}}{Q_i} \qquad (17)$$

However, calculation of the $Q$-factor from first principles is difficult because there are many different possible paths for mechanical energy dissipation. The most relevant mechanical loss sources in MEMS resonators are thermoelastic damping (TED), support losses, gas damping and damping due to surface imperfections, adsorbents or oxidation.

Duwel et al. [5] descibe a method to calculate the $Q$-factor as a result of TED using an eigenfrequency analysis. Their analysis shows that TED is an important loss mechanism for flexural modes. However comparison with measurements shows that for longitudinal and bulk modes other damping mechanisms seem to dominate. If resonators are operated at low pressures, gas damping is not limiting their $Q$-factor.

It seems that support losses are often the factor limiting the $Q$-factor of MEMS resonators. Support losses are caused by the generation of traveling waves, which transport energy from the resonator, via the supports and anchors, to the substrate. Because the waves are not localized near the resonator, their energy can essentially be considered as lost and is eventually converted to heat after being absorbed. The $Q$-factor of MEMS disk resonators as a result of support losses was both calculated analytically [6] and using a finite element method (FEM) calculation with perfectly matched layers (PML) [7].

## 7. Matched layers

The method we use to calculate the $Q$-factor as a result of support-losses using matched layers in COMSOL is similar to that presented by Bindel et al. [7]. Because it is unfeasible to model the full substrate, a model boundary is defined and it is assumed that the energy of all waves that cross this boundary are lost. To accomplish this absorption in a finite element calculation, matched layers are needed. Ideally these layers do not reflect acoustic waves and thus absorb all incident acoustic power. In the frequency response analysis mode of COMSOL 3.3a, perfectly matched layers (PML) are available for this purpose. However in the eigenfrequency analysis mode they are not implemented. Therefore we define simplified matched layers (ML). These ML layers are not as effective as PMLs because they only match perfectly for normal incident waves. However, as will be shown in section 10, $Q$-factors can be calculated within a few percent accuracy using such matched layers.

The matched layer consists of an artificial material with specifically chosen material parameters, such that it has a reflection coefficient $R=0$ for normal incident acoustic waves. Moreover, the acoustic waves that enter the ML, should be rapidly attenuated to ensure that they are absorbed before leaving the ML. This can be accomplished by a negative imaginary part of the wavevector in the ML (Im $k'<0$).

The reflection coefficient of an acoustic wave between two media is given by [4]:

$$R = \frac{Z'-Z}{Z'+Z} \quad (18)$$

Where $Z'$ is the acoustic impedance of the wave in the ML and $Z$ is the impedance in the isotropic medium from which the wave is incident. Clearly $Z'=Z$ is needed for zero reflection. For compressional waves $Z_{compr}=[\rho E(1-v)/((1+v)(1-2v))]^{1/2}$ and for shear waves $Z_{shear}=[\rho E/(2(1+v))]^{1/2}$, where $E$ is the Young's modulus and $v$ the Poisson ratio. The wave vector $k$ of the acoustic wave is given by

$$k = \omega\rho/Z \quad (19)$$

For a wave traveling through the ML in the positive $x$-direction the particle velocity is given by $v(x,t)=v_0 e^{j(\omega t-k'x)}$. For $k'=-j\alpha k$ the real part of the particle velocity will be proportional to $e^{-\alpha k x}$ and will attenuate exponentially as required (Re $\alpha>0$). The conditions $Z'=Z$ and $k'=-j\alpha k$ can be satisfied by defining the ML material properties $E', \rho'$ and $v'$ as:

$$E' = jE/\alpha$$
$$\rho' = -j\alpha\rho \quad (20)$$
$$v' = v$$

$E, \rho$ and $v$ are the material parameters of the material from which the acoustic wave is incident on the ML. The condition $Z'=Z$ is satisfied because $\rho E=\rho'E'$ and $v=v'$. $k'=-j\alpha k$ follows from equations (19) and (20).

Note that the substitution (20) attenuates waves in all directions and at all frequencies, for example a wave traveling through the ML in the negative-$x$ direction will be attenuated for increasing negative $x$.

As the ML works best for normal incident waves, the anchor is considered as a point source of acoustic waves and the ML boundary is oriented as a circular (2D) or spherical (3D) shell centered around the anchor. Moreover the ML should be thick enough to reduce the wave amplitude of the waves sufficient before they reflect from its outer boundary.

The ML can also be employed for wave absorption from anisotropic materials with an elasticity matrix **c**. In this case the ML should be given an anisotropic elasticity matrix **c'**$=j$**c**$/\alpha$ and $\rho'=-j\alpha\rho$. For normal incidence, the reflection $R$ is still zero and exponential attenuation of the wave in the ML still occurs. This can be understood from the fact that for any plane acoustic wave the acoustic impedance $Z$ does not change if $\rho$**c** stays constant, whereas the wavevector obeys $k=\omega\rho/Z$.

It is even possible to choose the material properties of the matched layer such that it will act as a PML [7,8]. Although this procedure is straightforward, it is much more elaborate than the transformation (20).

Because the material properties of the ML are complex-valued, the eigenvalues and eigenfrequencies $\omega_i$ will also become complex-valued, reflecting the exponential damping of the displacement field amplitudes in equations (2) and (10) with time in the absence of an excitation force $F_{ac}$. By substituting equation (2) in equation (16) we see that the total stored energy becomes time dependent:

$$E_{stored} = E_{stored,0}e^{-2t\operatorname{Im}\omega_i} \quad (21)$$

The $Q$-factor is defined as $2\pi$ times the ratio between the stored energy $E_{tot}$ and the dissipated energy per period $-dE_{stored}/dt*T$. The period $T=2\pi/Re(\omega_i)$. Therefore:

$$Q_i = \frac{E_{stored}}{-T\dfrac{dE_{stored}}{dt}} = \frac{\operatorname{Re}\omega_i}{2\operatorname{Im}\omega_i} \quad (22)$$

Using equation (17) we can thus calculate $b_i$.

The equations in sections 4 and 5 remain valid for damped systems with complex eigenvectors and eigenvalues within the limits discussed in [3]. Because the absolute signs in equation (16) are outside the integral, the energy of traveling waves is not included in the stored energy integrals.

## 8. Application example

As an example of the method described above, the parameter extraction for the MEMS radial disk resonator in figure 1 will be performed. In figure 3 the layout of the resonator and its substrate is given for $R=11$ μm, $H=3$ μm, $h=0.8$ μm and $a=0.8$ μm. The radius of the silicon substrate is 20 μm and the outer radius of the ML is 40 μm. The materials are assumed to be isotropic. The material parameters of the disk, stem and substrate are respectively given by: $E_d=1061$ GPa, $\rho_d=3440$ kg/m$^3$, $v_d=0.12$, $E_s=150$ GPa, $\rho_s=2330$ kg/m$^3$, $v_s=0.22$, $E_s=130$ GPa, $\rho_s=2330$ kg/m$^3$, $v_s=0.28$. The material parameters of the matched layer are determined by applying transformation (20) on the substrate parameters, with $\alpha=3$. The mapped mesh consists of 3168 quadrilateral quadratic Lagrange elements.

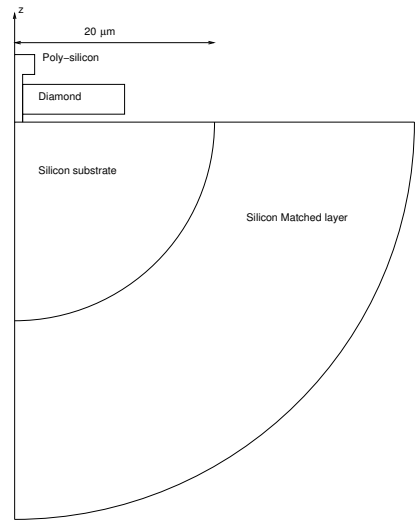

**Figure 3** Layout of the simulated diamond disk resonator with axial symmetry around the $z$-axis.

By solving the eigenfrequency problem, the first radial bulk-mode is found at 489.27 MHz. The total displacement mode-shape of the resonator is shown in figure 4. The extension of the disk creates stress in the stem, which generates acoustic waves in the substrate. In figure 5 the same mode-shape is plotted over a smaller displacement range. Spherical substrate waves are generated by the stem of the resonator. A variation of the phase of the solution shows that these traveling waves propagate outward and are absorbed by the ML. Inside the matched layer the wave amplitude decays exponentially as expected.

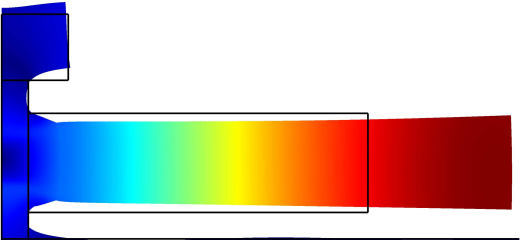

**Figure 4** Total displacement in the first radial bulk-mode of the disk resonator at 489.27 MHz.

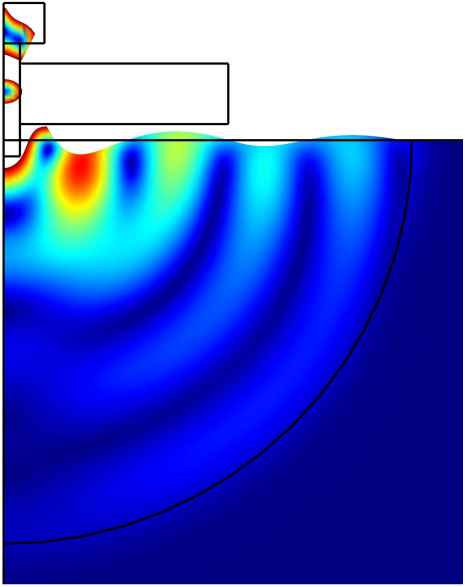

**Figure 5** Acoustic substrate waves generated by the first radial bulk-mode of the disk resonator.

## 9. *Q*-factor comparison

With a script, the *Q*-factor of the first radial bulk-mode of the diamond disk resonator is calculated from the complex eigenfrequency using equation (22) for 56 different disk radii *R* between 7.5 and 13 μm. The results from this FEM calculation are shown in figure 6. In the same figure the measured *Q*-factors from [1] and the analytical calculated *Q*s from [6] are plotted for *R*=8,10,11 and 12 μm. The FEM simulations show a good agreement with the analytic results from [6]. The difference with the measured *Q*s is larger, this can be due to process uncertainties which result in slight deviations of the actual stem size, as was also discussed in [6]. The lines which represent the FEM simulations in figure 6 show a slight periodic variation in *Q* as a function of frequency. The period of this variation depends on the radius of the silicon substrate in figure 3. Therefore the variation is attributed to a small mismatch between the substrate and the ML, which will affect the effective acoustic impedance presented by the substrate to the resonator.

The *Q*-factor for the second radial bulk-mode is also calculated. For *R*=11 μm and *a*=0.8 μm we find for the second radial bulk-mode *Q*=1,900, which is about a factor 5 lower than the measured and calculated *Q*-factors of around 10,000 in [1] and [6]. It is unclear what the cause is for this large difference.

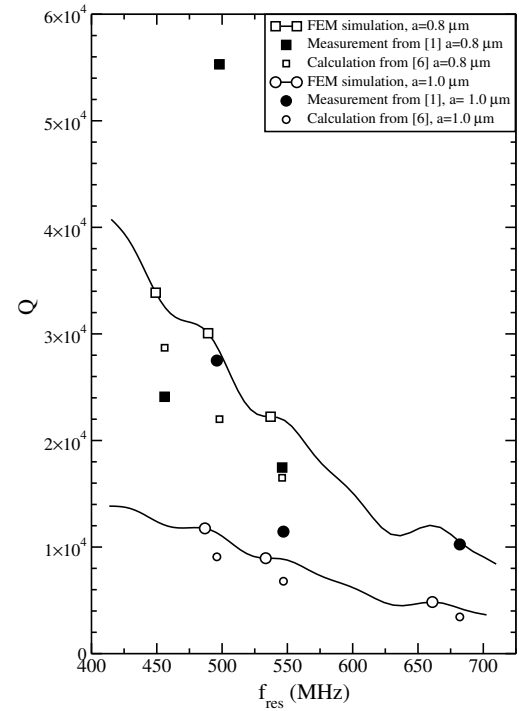

**Figure 6** *Q*-factor and resonance frequency of the first radial bulk-mode of the diamond disk resonator with radius *R* between 7.5 and 13 μm. FEM simulations (solid lines+symbols) from this work are compared to the results from [1] and [6] for *R*=8,10,11 and 12 μm.

## 10. Frequency response and PML

To compare the results obtained from the eigenfrequency analysis to a frequency response analysis, the same structure with ML described in section 8 is analyzed in the frequency response mode. Using equation (14) an actuation pressure $V_{ac}\eta/A_{act}$ is applied to the side of the disk in the *r*-direction. The displacement $x_i$ as a function of frequency is determined using equation (13). From equation (6) it is seen that $i_{ac}=j\eta\omega x_i$. Thus the admittance is determined from the frequency response mode using

$Y(\omega)=i_{ac}/V_{ac}=j\eta\omega x_i/V_{ac}$. The same analysis is also performed with the ML replaced by a cylindrical PML.

These frequency response results are compared to the calculation of $Y(\omega)$ using equation (8), where $k_i, m_i$ and $b_i$ are calculated using equations (16) and (17). $C_w$ is kept to be zero. The three admittance curves are shown in figure 7. The agreement between the calculation in eigenfrequency mode and frequency response mode is excellent, which confirms the correctness of the described procedure for calculating the admittance from the eigenmodes and eigenfrequencies. Analyzing the admittance curves we find that the $Q$-factor determined with the ML is 30,068 whereas the $Q$-factor as determined with the PML is 29,006. The lost energy is proportional to $1/Q$. If we assume that the PML is ideal we can thus conclude that the ML absorbs $Q_{PML}/Q_{ML}$=96.5% of the incident energy. If more accurate values for the $Q$-factor are needed it is to be recommended to use PML boundaries. This is especially true if the acoustic waves are incident on the ML surface at more grazing angles than in this example.

Comparison of the simulation times of the eigenfrequency analysis and frequency response analysis shows that it takes a factor 60 less simulation time to calculate 1 eigenvalue than to perform a frequency response analysis at 100 frequencies. Even if we consider that the shape of the curve might be determined using less frequencies, the eigenfrequency analysis will run in a significantly shorter simulation time.

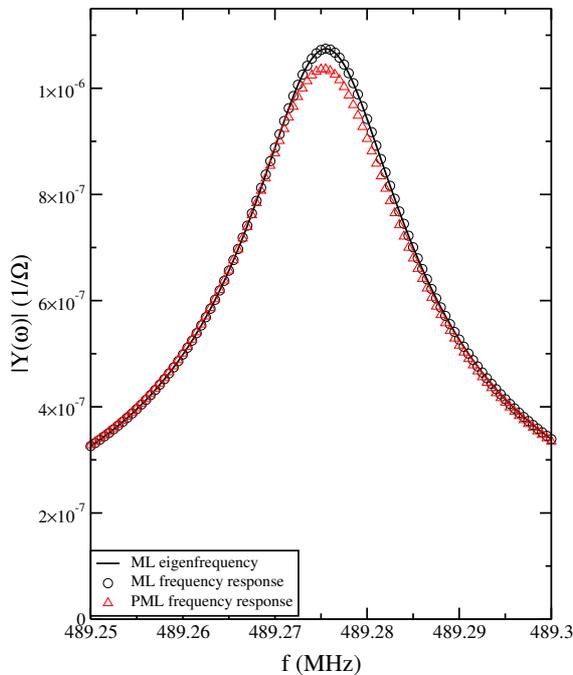

**Figure 7** Admittance of the first radial bulk-mode of the diamond disk resonator with $R$=11 μm and $a$=0.8 μm. Equation (8) with ML is compared to the frequency response mode analysis with ML and PML. $V_{dc}$=2.5V, $g$=90 nm [1].

## 11. Conclusions

This paper presents a simulation method to determine the frequency dependent admittance of a MEMS resonator by postprocessing of the solutions a FEM eigenfrequency simulation. Support-losses are modeled using a matched layer (ML) to represent the absorption of acoustic waves in the substrate. The method is applied to a MEMS diamond disk resonator with varying radius and stem diameter. Calculated $Q$-factors agree well with measured and analytic values. A comparison with a frequency response analysis confirms the validity of the method and shows that the ML absorbs about 96.5% of the incident acoustic energy. Moreover it shows that the eigenfrequency analysis takes considerably less simulation time.

## 12. References

1. J. Wang et al., 1.51-GHz Nanocrystalline Diamond Micromechanical Disk Resonator With Material-Mismatched Isolating Support, *Proc. MEMS 2004*, pp. 641-644

2. H.A.C. Tilmans, Equivalent circuit representation of electromechanical transducers I&II, *J. Micromech. Microeng.*, **6**, p.157-176 (1996), *ibid.* **7**, p.285-309 (1997)

3. W. Weaver, Jr., S.P. Timoshenko, D.H. Young, *Vibration problems in engineering*, 5$^{th}$ ed., p. 275. John Wiley&Sons, New York (1990)

4. B.A. Auld, *Acoustic fields and waves in solids Vol. I*, p. 42. John Wiley&Sons, New York (1973)

5. A. Duwel et al., Engineering MEMS Resonators With Low Thermoelastic Damping, *J. MEMS*, **15** (6), p. 1437-1445 (2006)

6. Z. Hao and F. Ayazi, Support loss in the radial bulk-mode vibrations of center-supported micromechanical disk resonators, *Sensors and Actuators A*, **134**, p. 582-593 (2007)

7. D.S. Bindel and S. Govindjee, Elastic PMLs for resonator anchor loss simulation, *Int. J. Numer. Meth. Engrg.*, **64**, pp. 789–818 (2005)

8. U. Basu and A.K. Chopra, Perfectly matched layers for time-harmonic elastodynamics of unbounded domains: theory and finite-element implementation, *Comput. Methods Appl. Mech. Engrg.* **192,** pp. 1337-1375 (2003)